\pdfoutput=1

\documentclass[11pt]{article}
\usepackage{jheppub}

\usepackage{amsmath}
\usepackage{amssymb}
\usepackage{graphicx,color,slashed,braket}
\usepackage{bbm}
\usepackage{ifpdf}
\usepackage{comment}

\usepackage{booktabs}
\usepackage{multirow}
\usepackage{makecell}
\usepackage{float}
\usepackage{verbatim}
\usepackage{caption}
\usepackage{cancel}
\usepackage{enumerate}
\usepackage{tcolorbox}
\usepackage[font=small,labelfont=bf]{caption}

  \usepackage{color}
  \definecolor{dark-gray}{gray}{0.20}
  \definecolor{gray}{gray}{0.30}
  \definecolor{light-gray}{gray}{0.80}
  \definecolor{dark-red}{rgb}{0.7,0,0}
  \definecolor{dark-green}{rgb}{0.1,0.4,0}
  \definecolor{dark-blue}{rgb}{0.3,0.3,0.7}
  \definecolor{light-blue}{rgb}{0.8,0.8,1}

\usepackage{hyperref}
\usepackage{setspace}
\hypersetup{
	colorlinks=true,
	linkcolor=dark-blue,
	citecolor=dark-red,
	urlcolor=dark-green,
	linktoc=page
}
\newcommand{\be}{\begin{equation}}
\newcommand{\ee}{\end{equation}}

\def\be{\begin{equation}}
\def\ee{\end{equation}}
\def\bea{\begin{eqnarray}}
\def\eea{\end{eqnarray}}

\newcommand{\vol}{\text{vol}}

\newcommand{\Mpl}{M_{\textrm{Pl}}}

\newcommand{\dd}{\mathrm{d}}

\allowdisplaybreaks

\title{Over-extremal brane shells from string theory?}

\author{Ulf Danielsson$^1$, Vincent Van Hemelryck$^2$, Thomas Van Riet$^2$}

\affiliation{$^1$
Institutionen f\"or Fysik och Astronomi,
Box 803, SE-751 08 Uppsala, Sweden\\
$^2$
Instituut voor Theoretische Fysica, K.U.Leuven,
Celestijnenlaan 200D, B-3001 Leuven, Belgium}
\emailAdd{ulf.danielsson@physics.uu.se}
\emailAdd{vincent.vanhemelryck@kuleuven.be}
\emailAdd{thomas.vanriet@kuleuven.be}

\notoc

\abstract{We demonstrate that, if the usual phenomenological  compactifications of IIB string theory with warped throats and anti-branes make sense, there must exist spherical brane shells in 4d that are overcharged. They correspond to classical over-extremal objects but without the usual naked singularities. The objects are made from D3-particles that puff into spherical 5-branes that stabilise at finite radii in 4d  and whose inside corresponds to the supersymmetric AdS vacuum. One can think of these shells as stabilised Brown-Teitelboim bubbles. We find that these objects can be significantly larger than the string scale depending on the details of the warped compactification. }

\begin{document}
 \begin{flushright}
 UUITP-26/22 
 \end{flushright} 
\maketitle

\newpage

\section{Introduction}
String theory has shown to provide the rich structure needed to describe the quantum nature of black holes, at least for black holes that are supersymmetric or are close to being so. For charged non-rotating black holes this means they are necessarily extremal. The main success of string theory has been in counting the microstates \cite{Sen:1995in, Strominger:1996sh}, but a clear understanding of the spacetime structure of the inside of a black hole is yet to be found, especially for black holes that are more realistic in an astrophysical sense and hence are not extremal. Many bottom-up ideas exist for what could replace the interior of a black hole, but for actual top-down constructions it is less clear. Aside the fuzzball proposal \cite{Mathur:2005zp, Bena:2022rna} there is another appealing possibility as first described in \cite{Danielsson:2017riq}: black holes could be spherical shells whose inside correspond to a stable supersymmetric AdS space whereas the outside is a meta-stable vacuum. When the meta-stable vacuum decays through bubble nucleation, the bubbles tend to expand and overtake the universe, unless there is a mechanism that stabilises them so they remain at finite radius. When they do so, the objects behave in many ways as sensible black holes to an outside observer \cite{Danielsson:2017riq,Danielsson:2017pvl, Danielsson:2021ruf, Danielsson:2021ykm}. Although this scenario is highly ``string-inspired'', an actual embedding within string theory is lacking. In particular one needs to find the mechanism that stabilises the radius of the bubble. It is the aim of this paper to provide such an embedding, but we will see that our examples are actually describing over-extremal objects. This by itself is very intriguing since it is highly unclear whether over-extremal objects, like the charged particles of the Standard Model, can be macroscopic in size as well. Although our objects are over-extremal and so not of astrophysical interest, if the working assumption in this paper is correct (ie phenomenological IIB compactifications) then our results show that at least the principle of stabilising bubbles with SUSY AdS vacua inside does occur within string theory. Whether the size of our over-extremal balloon like objects can truly become large (beyond particle physics length scales or the de Broglie wavelength) seems to depend on how far flux numbers, warping effects and spacing between vacuum energies can be tuned. The possibility of this tuning is a crucial debate within the current Swampland paradigm \cite{Vafa:2005ui, Brennan:2017rbf, Palti:2019pca} that has been discussed for instance in \cite{Bena:2018fqc, Gao:2020xqh, Carta:2021lqg, Plauschinn:2021hkp, Marchesano:2021gyv, Bena:2020xrh,  Grana:2022dfw, Broeckel:2021uty, Bastian:2021hpc, Demirtas:2021nlu, Demirtas:2021ote, Crino:2022zjk}.

It is well-known in string theory that brane bubbles can be stabilised at finite radii, see e.g. the mechanism for brane-flux decay \cite{Myers:1999ps, Kachru:2002gs, Gautason:2015tla} but in all such examples the bubbles form inside the extra dimensions. To our knowledge we provide the first mechanism where this happens inside the external dimensions. The ideas for our construction originated from discussions in reference \cite{Montero:2021otb}.

The rest of this paper is organised as follows. In section \ref{sec:D3} we explain the set-up of wrapped D3-particles in the context of uplifting scenario's involving space-filling anti-D3-branes and we explain how to compute the potential energy as a function of the radius of the puffed-up D3-particles into spherical 5-branes. In section \ref{sec:shells} we then investigate in detail the stabilised brane shells that emerge: their size, mass and charge. We end with a discussion in section \ref{sec:disc}.

\section{Puffed up and wrapped D3-branes }\label{sec:D3}
One way to construct massive point-like objects from 10-dimensional string theory is to wrap D$p$ branes over $p$-dimensional cycles inside the extra dimensions. If this can be done in a stable manner, one finds a particle in 4d. If we wrap many branes the particle can be more massive and even become black-hole like.

Let us take the example of IIB string theory and wrap D3-branes over 3-cycles. The resulting particles will be charged under a U(1) gauge field $A_{\mu}$ that comes from reducing the RR gauge potential $C_4$ over that 3-cycle with volume tensor $\epsilon_3$:
\be
C_4=A_{\mu}dx^{\mu}\wedge\epsilon_3+\ldots
\ee
where $x^{\mu}$ are 4d coordinates. Intuitively one can think of this charge as the reason the particle is stable since it will be BPS once the cycles is supersymmetrically calibrated. In other words, the charge guarantees the stability by means of the BPS bound. However, for compactifications with less than $\mathcal{N}=2$ supersymmetry, the situation is more involved. For example, let us consider IIB reduced over Calabi-Yau orientifolds, the starting point for the majority of string-phenomenological moduli stabilisation scenarios. In this case the orientifold projections make it impossible for D3-particle states to be stable \cite{EnriquezRojo:2020hzi}, essentially because the gauge fields are projected out by the orientifolding.

In what follows we construct massive D3-states on orientifold-odd cycles, and give them a charge by relying on the Freed-Witten effect and the presence of anti-branes in uplifting scenarios, inspired from the discussions in \cite{Montero:2021otb}. In simple words: adding anti-D3-branes to uplift the KKLT \cite{Kachru:2003aw} or the LVS scenario \cite{Balasubramanian:2005zx} AdS vacuum\footnote{See \cite{Danielsson:2018ztv, Gao:2020xqh, Lust:2022lfc, Junghans:2022exo, DallAgata:2022abm, Gao:2022uop} for some recent works discussing the very existence of these vacua.}  introduces exactly the U(1) gauge field to charge the objects, which we will see is really necessary for our purposes. 

The detailed set-up is the following: consider D3-branes wrapping the A-cycle of the KS throat \cite{Klebanov:2000hb} embedded in some compact Calabi-Yau geometry. Since there are $M$ units of $F_3$ piercing that cycle as well, there are $M$ fundamental
strings leaving from every D3-brane that attach to the anti-D3-branes, by means of the Freed-Witten effect \cite{Freed:1999vc}. 

In what follows, we show that the corresponding D3-particles polarise into a spherical NS5-brane shell. We first discuss this in the usual way with probe actions and then use GR techniques to compute backreaction effects. We will find that qualitatively there is not much difference, but the comparison shows nicely how the more bottom-up ideas from \cite{Danielsson:2017riq}, using junctions conditions, are indeed realised in string theory.

\subsection{Brane polarisation in 4d using probes}

There are two ways to investigate whether D$p$-branes puff into spherical branes D$(p+2)$ \cite{Myers:1999ps}. Either using the non-Abelian worldvolume theory on the D$p$-brane or using the Abelian D$(p+2)$ action. We will follow the second approach. This means that we consider a NS5-brane of spherical shape in 4d that, when it pinches to a point, describes the worldvolume action of a D3-particle charged under the space-filling anti-D3 worldvolume gauge field. Then we compute the energy in terms of its radius and check whether it has a meta-stable minimum at finite radius. There will be three contributions to the energy of the brane shell: the DBI energy, the Coulomb energy and the pressure energy. The latter is related to the fact that the inside of the bubble is SUSY AdS and the outside is essentially Minkowski (de Sitter with small vacuum energy). Let us compute these 3 separate ingredients:\\

{\bf The DBI energy}:  The DBI action of the NS5-brane, in string units, is 
\be
    S_\text{NS5} = -\frac{\mu_5}{g_s^2} \int \dd x^6 \sqrt{-\det(g_{MN} + 2\pi  g_s \mathcal{F}_{MN})}\,,
\ee
with $2\pi \mathcal{F}_2 = 2\pi F_2 - C_2$. The WZ part of the NS5 action will be specified later.  The DBI action of a D3-brane instead goes like
\be
S_\text{D3} = -\frac{\mu_3}{g_s} \int \dd x^4 \sqrt{-\det(g_{ab}+\ldots)}
\ee
where, in string units, $4\pi^2\mu_5 = \mu_3$.  In order to compute the potential energy as a function of the radius we need to know the induced metric, the worldvolume flux $F_2$ and the RR potential $C_2$.   If we call the radius of the NS5-brane $R$ then the string frame metric induced on the worldvolume is:
\begin{align}
    \frac{\dd s_6^2}{\ell^{2}_s} =  e^{2A_0}\left(- \dd T^2 + R^2 (\dd \theta^2 + \sin^2 \theta \dd \phi^2) \right)  + b_0^2 g_s M \left(\dd \psi^2 + \sin^2 \psi ( \dd \chi^2 + \sin^2 \chi \dd \xi^2) \right)\,,
\end{align}
where we used a standard expression for the metric at the tip of the Klebanov-Strassler throat \cite{Kachru:2002gs} (using that the B-cycle pinches. The warpfactor at the tip is denoted $e^{2A_0}$, $M$ is the $F_3$-flux quantum through the A-cycle, $g_s$ is the string coupling and $b_0^2\approx 0.93$ an order one numerical constant. The coordinates $t, \theta$ and $\phi$ are in the external space and the coordinates $\psi, \chi, \zeta$ parametrise the local A-cycle of the internal compact dimensions. From here onward we work in string units where $\ell_s=1$.

In order for the spherical NS5 to correspond to $N_\text{D3}$ D3-branes when $R=0$, the NS5 world volume flux should be:
\be 
    F_2 = \frac{N_\text{D3}}{2} \vol(S^2) =  \frac{N_\text{D3}}{2} \sin \theta \dd \theta \wedge \dd \phi\,,\qquad \frac{1}{2\pi} \int_{S^2} F_2 = N_\text{D3}\,.
\ee
To find the potential we integrate the DBI term over all spatial coordinates and use that $S_\text{DBI}(R, \dot{R}=0) = -\int \dd T\: V_\text{DBI}(R)$. This will be interpreted as the gravitational term responsible for the shell wanting to collapse under its own weight.  With the background expression of $C_2$ near the tip, 
\be
    C_2 = M \left(\psi -\frac{1}{2} \sin(2 \psi) \right) \sin \chi \dd \chi \wedge \dd \xi\,,\qquad      \int_{S^3} \dd C_2 = (2\pi)^2 M\,,
\ee
the DBI potential becomes 
\be
    V_\text{DBI}(R) = \mu_3 \left( 4\pi \tau e^{A_0} \sqrt{(\pi g_sN_{\text{D3}})^2 + (e^{A_0}R)^4}\right).
\ee
The tension-like parameter $\tau$ is defined as 
\begin{align}\label{eq:tau}
\tau = 2\frac{(g_s M)^{\frac{3}{2}}}{g_s^2} \mathcal{C}\,,
\end{align}
with $\mathcal{C}$ the following numerical constant
\be \label{eq:C}
    \mathcal{C} = b_0  \int \dd \psi \sqrt{4 b_0^4 \sin^4 \psi + (-2 \psi + \sin 2 \psi)^2} = 10.3591\,.
\ee

{\bf The Coulomb energy}: The shell must be charged under the gauge theory coming from the anti-D3s due to the Freed-Witten effect. We implement this by adding the following coupling.
\be
    S_\text{Coulomb} = \alpha\mu_5 \int_{\Sigma_\text{NS5}} F_3 \wedge \tilde{A}_1 \wedge F_2,
\ee
where $\alpha$ is a constant and $\tilde{A}_1$ is the worldvolume gauge field of the anti-D3.
These terms we added by hand since there is no supersymmetric brane action anymore to guide us. Nevertheless this coupling seems quite unique and when integrating it over all spatial coordinates we get the expected potential for a static charged shell:
\be
    V_\text{Coulomb} =  \alpha\mu_3 \frac{g^2Q^2_\text{shell}}{2R}\,,
\ee
Where $Q_\text{shell} = M N_\text{D3}$ is the charge of the shell and $g^2 = \pi g_s$ is the gauge coupling. This is to be compared with the textbook expression for the energy of a charged shell
\be
    V_\text{Coulomb} =  \frac{1}{4\pi}\frac{g^2Q^2_\text{shell}}{2R}\,,
\ee
Since we worked in string units where $\ell_s=1$ and we have $\mu_3 = (2\pi)/(2\pi \ell_s)^4$, this fixes $\alpha$ to be $\alpha = 2\pi^2 $. \\

{\bf The pressure energy}:  Since the D3-shell is understood as an NS5-brane, it is expected that the $H$-flux on the B-cycle drops by one unit, $K \to K-1$, i.e. it triggers a brane/flux decay, as one passes the shell. Therefore the vacuum energy density inside the charged sphere will be lower than outside.  If we assume that the energy density difference between the uplifted and SUSY vacuum is $2 \mu_3 e^{4A}$, we have a pressure energy contribution to the potential: 
\be
    V_\text{pressure } = -  2 g_s^{-1}\mu_3 e^{4A} \frac{4\pi}{3} R^3\,.
\ee

{\bf The complete potential}: The total potential $V$ becomes $V=V_\text{DBI} + V_{\text{Coulomb}} + V_{\text{pressure}}$:
\be \label{eq:stringpot}
 V = e^{A_0}\mu_3  \left(4\pi \tau\sqrt{\left(e^{A_0} R\right)^4 + N_{D3}^2\pi^2 g_s^2}+  2 \pi^3 g_s\frac{(M N_\text{D3})^2}{2 (e^{A_0}R)} -  2\frac{4\pi}{3} g_s^{-1}(e^{A_0}R)^3\right)\,.
\ee
We see that the potential only depends on the combination $e^{A_0}R$. Hence the potential stabilises $e^{A_0}R$ at a value that depends on the coefficient in $V/e^{A_0}$, which are \emph{warping independent}. So the size of $e^{A_0}R$  depends on $M,N_{\text{D3}}$ and $g_s$. Note that the radius measured by the 10d metric
$\dd s^2_{10}=e^{2A_0} \dd s_4^2 +\ldots$ is shrunk by a factor $e^{A_0}$ and so the physical size of the shell in string units is $R_\text{phys}= Re^{A}$.

\subsection{Brane polarisation beyond probe level using junction conditions}
Using Smarr relations and blackfold techniques, brane polarisation has been discussed beyond probe level in for instance \cite{Armas:2018rsy,Armas:2019asf,Nguyen:2021srl,Cohen-Maldonado:2015ssa,Cohen-Maldonado:2016cjh}. The more pedestrian 4d GR-inspired method uses Israel junction conditions instead as in the original proposal of \cite{Danielsson:2017riq}. In here we merge the probe results of the previous section with the junction conditions and find corrections to the probe expressions, which however do not alter the final results much.\footnote{In \cite{Danielsson:2022fhd}, the relation between the brane action and junction conditions was explored in a different context.} This serves as a very non-trivial consistency check of the brane potential and provides a stringy interpretation to the bottom-up procedure of junction conditions.

We start out with a charged shell in 4d with a Reissner-Nordström metric on the outside and an AdS metric on the inside. In principle the spacetime on the outside should be asymptotically (A)dS but since the cosmological constant is supposed to be tiny, it is safe to ignore it. We therefore work with the ordinary RN-metric which is asymptotically Minkowski.
Both the RN (+) and AdS (-) metrics take the following form
\be
    \dd s_{\pm}^2 = -f_\pm(r) \dd t^2 + \frac{\dd r^2}{f_{\pm}(r)} + r^2 \dd \Omega_2^2,
\ee
where the $f_\pm(R)$ are
\begin{align}
    & f_+(r) = 1 - \frac{2G_4 m}{r} + \frac{G_4(gQ)^2}{4\pi r^2} \,,\\ 
    & f_-(r) = 1 + k^2 r^2\,.
\end{align}
Note that now $r, t$ are now dimensionful, unlike $R, T$ before.
The shell that separates the RN exterior from the AdS interior is in our case a wrapped NS5-brane. The junction condition across this shell is obtained by setting the difference of the extrinsic curvature equal to the effective tension $\mathcal{T}$ of the shell. The Israel junction condition then becomes \cite{Israel:1966rt}
\be\label{JUNCTION}
    \mathcal{T}(r) =  \frac{2}{8\pi G_4}\frac{1}{r}\left(\sqrt{f_-(r)}-\sqrt{f_+(r)}\right)\,.
\ee
The benefit of having a top down description is that we can fix the effective tension from the previous subsection. The effective tension $\mathcal{T}(r)$ depends on the radius of the shells and can be found from the DBI part of the NS5 action as follows
\be
S^\text{DBI}_\text{NS5} =  \mathcal{T}(r)\int \sqrt{-h}\, \dd^3\,x\,.
\ee
with $h$ the induced metric on the 3d shell. To compare the junction condition analysis with the previous section we need to compare coordinates. In order to compare them note that the previous section ignored backreaction, which means $f_{\pm}=1$.  Hence we have
\be
t = \ell_s e^{A_0} T\,,\qquad r = \ell_s e^{A_0} R\,.
\ee
To find the tension we use (from equating the on-shell actions) that
\be
4\pi\mathcal{T}r^2\dd t = V^{\text{DBI}}_\text{NS5} \dd T\,,
\ee
which implies:
\be
    \mathcal{T}(r) = \frac{\mu_3 \ell_s \tau} {r^2} \sqrt{(\pi g_sN_{\text{D3}})^2\ell_s^4 + r^4}.
\ee
Note that in the large $r$ regime the tension becomes $r$-independent which is the usual assumption in the bottom-up literature.
Combining everything together, one can rewrite the junction condition as:
\be\label{JUNCTIONPLANCK}
    \mu_3\ell_s\tau   =  \frac{2}{8 \pi G_4} \frac{r}{\sqrt{(\pi g_sN_{\text{D3}})^2\ell_s^4 + r^4}} \left( \sqrt{1+ k^2 r^2 } - \sqrt{1- \frac{2G_4 m}{r} + \frac{G_4 (gQ)^2/4\pi}{r^2}}\right)\,.
\ee
The shell potential can then be obtained by multiplying the tension with the square root determinant of the induced metric, which is done such that one takes the averaged determinant roots. The other pieces in the potential are argued in the usual way and coincide with the probe expressions. In the end we obtain:
\begin{align}
    V =& 4\pi r^2 \mathcal{T}(r)\frac{1}{2} \left(\sqrt{1 + k^2 r^2} + \sqrt{1 - \frac{2G_4 m}{r} + \frac{G_4 (gQ)^2/4\pi}{r^2}} \right)\nonumber\\ 
    &- \frac{1}{2 G_4} \left( k^2 r^3 - \frac{G_4 (gQ)^2/4\pi}{r}\right) \,.\label{VJUNCTION}
\end{align}
The consistency of this procedure can easily be verified by using the junction condition to write the tension as in the right hand side of equation \eqref{JUNCTION} such that all terms drop against each other and we find the following on-shell value of the potential:
\be
V= m\,,
\ee
which indeed is the total energy (mass). The difference between the approach based on the junction condition and the brane probes can then be understood as follows; in the junction condition formalism one has an extra parameter in the potential, namely $m$, but one also has an extra condition: namely $V$ at the minimum has to equal this parameter $m$. Whereas in the probe approach one simply minimises the potential and $m$ never appeared as an external parameter. The junction condition formalism can be seen as corrected version of the brane probe formalism, but for our purposes the two approaches give qualitatively similar results.

To compare \eqref{VJUNCTION} with (\ref{eq:stringpot}), we use\footnote{The absence of warping in the last equation is due to a rescaling of coordinates.}
\begin{align}
 Q=MN_\text{D3}\,,\qquad  g^2=\pi g_s\,,\qquad 3k^2\Mpl^2 = \frac{2}{g_s}\mu_3 \,.
\end{align}
The only difference is that the junction conditions take into account backreaction on the bulk metric such that $\mathcal{T}(r)r^2$ is now not multiplied with a constant, but with an $r$-dependent function which becomes $1$ in the probe approximation.

\section{The stabilised brane shells}\label{sec:shells}

The shells have a radius that is determined by minimising the potential \eqref{VJUNCTION}.
In order to proceed we will work in Planck-like units where $G_4 = 1$ and ignore the cosmological constant assuming that the Hubble length is much larger than the shell radius. The potential to minimise then becomes:
\begin{align}
    V &= \frac{ \ell_s^{-3} \tau}{(2\pi)^2}\sqrt{(\pi g_s N_\text{D3})^2\ell_s^4+ r^4} \left(1 + \sqrt{1 - \frac{2 m}{r}  + \frac{ (gQ)^2/4\pi}{r^2}} \right) +  \frac{(gQ)^2/4\pi}{2r} \,. \label{VPLANCK}
\end{align}
Since the D3-brane tension is $\mu_3 = 2\pi/(2\pi \ell_s)^4$, we need to express the string length $\ell_s$ in Planck units, which should be a large number.

In these units, the junction condition \eqref{JUNCTION} becomes
\be\label{JUNCTIONPLANCK2}
    \frac{1}{\pi^3\ell_s^3}  \tau = \frac{2 r}{\sqrt{(\pi g_s N_\text{D3})^2\ell_s^4 + r^4}} \left(1- \sqrt{1-\frac{2m}{r}+ \frac{(gQ)^2/4\pi}{r^2}} \right)\,.
\ee

\begin{figure}[h]
    \centering
    \includegraphics[width=0.5\textwidth]{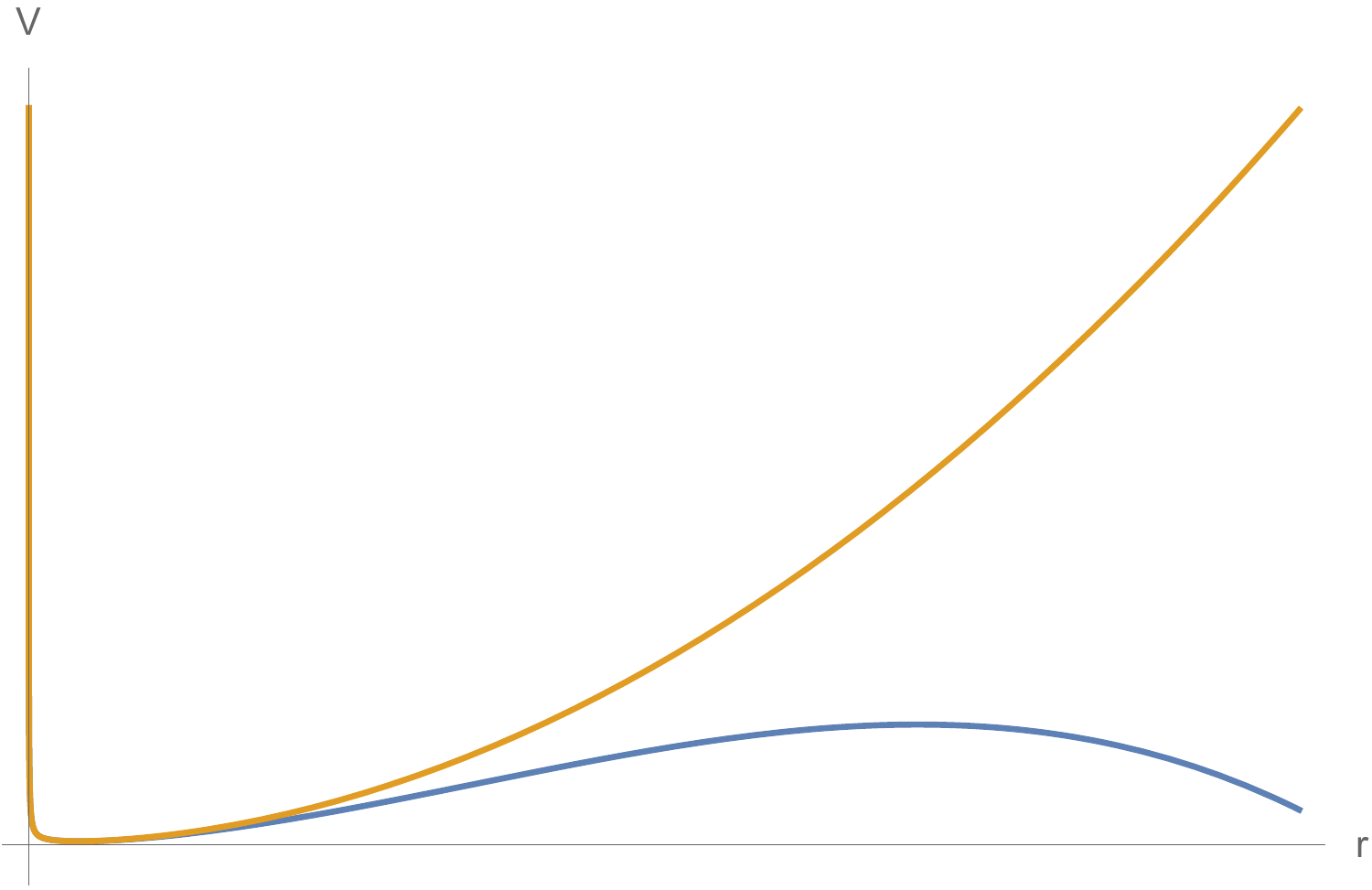}
    \caption{A sketch of the potential $V$ as a function of $r$, with and without the cosmological constant. The blue curve includes the cosmological contribution and the orange is one without. For the sake of illustration we have plotted the potential for a large cosmological constant. }
    \label{fig:plotV}
\end{figure}

To highlight our approximation it is informative to make a plot of the potential in \eqref{VPLANCK} versus the potential of \eqref{VJUNCTION}. In figure \ref{fig:plotV} we have plotted both on top of each other, where the blue curve has the cosmological constant included and the orange curve is without. The orange curve only has one minimum corresponding to the brane shell whereas the blue curve also has a maximum and then a runaway direction. In the limit of small cosmological constant this runaway piece and the maximum will be send to very far distances, corresponding to the fact that the nucleation probability toward the true vacuum is highly suppressed. More accurately our assumptions are that the length scale associated to the \emph{difference} in vacuum energies (inside and outside the bubble) is parametrically larger than the radius of the bubble.  For jumps in the cosmological constant  that are not small, one can verify that it becomes more difficult to find a meta-stable brane shell.

From now onward we therefore continue with the approximation that the cosmological constant can safely be ignored and use \eqref{VPLANCK}.  In what follows we will discuss the minima of this potential.
Interestingly it will turn out that minimising the potential at finite radius is never possible for extremal or sub-extremal configurations.  

In order to find minima of the potential we use that 
\be
gQ/\sqrt{4\pi} = \frac{\sqrt{g_s}}{2}M N_\text{D3}>1\,.
\ee
This can be argued from  $g_s M \gg 1$ such that the A-cycle can be described within supergravity. Alternatively this condition can be argued from the stability of the conifold modulus \cite{Bena:2018fqc}.

From equation \eqref{JUNCTIONPLANCK2} one can infer a maximal value of $\tau$. This works as follows. The only negative contribution to $\tau$ comes from the square root, which would vanish at the horizon radius. The overall term is largest for minimal $r$ and a minimal horizon radius occurs for an extremal black hole. Hence $\tau$ is maximised for extremal configurations where $m= gQ/\sqrt{4\pi}$ at the extremal radius $r=m$: 
\be\label{taumax}
    8\pi \mu_3 \tau< 8\pi \mu_3 \ell_s \tau_\text{max} = \frac{2 gQ/\sqrt{4\pi}}{\sqrt{(\pi g_s N_\text{D3})^2 \ell_s^4+(gQ)^4/(4\pi)^2}}.
\ee
This bound can be translated into a bound for our string parameters, and is given by 
\begin{align}\label{taumax2}
    \frac{2\mathcal{C}}{ (2\pi)^2} (\ell_s^{-2}M)^{\frac{1}{2}}(\ell_s^2 N_\text{D3})^{-1} \sqrt{16\pi^2 (\ell_s^2 N_\text{D3})^2 + (\ell_s^{-2}M)^4 (\ell_s^2 N_\text{D3})^4} <1.
\end{align}
Recall that $\mathcal{C}$ is a numerical constant defined in \eqref{eq:C} and $\ell_s$ here is a number since it is the string length in Planck units.  Remarkably this bound does not depend on the string coupling and only involves the combinations $\ell_s^2 N_\text{D3}$ and $\ell_s^{-2} M$.

We are now ready to numerically determine the shell radius $r_*$ and mass $m$ as a function of $M, g_s, \ell_s, N_{\text{D3}}$ from the equations:
\begin{equation}
\label{eq:StableShellConditions}
    \partial_r V(r)|_{r=r_*} =0, \qquad V(r_*) = m.
\end{equation}
To solve the system, it is useful to rewrite $r$ and $m$ as $r \equiv \frac{gQ}{\sqrt{4\pi}} \,\delta$ and $m \equiv \frac{gQ}{\sqrt{4\pi}}\, \epsilon$, hence over-extremality means that $\epsilon<1$. The potential then depends trivially on the string coupling:
\begin{align}
   V &= \sqrt{g_s}\left[\frac{\ell_s^{-3} M^{3/2}\mathcal{C}}{2(2\pi)^2}\sqrt{(4\pi N_\text{D3})^2\ell_s^4+ (M N_\text{D3})^4} \left(1 + \sqrt{1 - \frac{2 \epsilon}{\delta}  + \frac{1}{\delta^2}} \right) +  \frac{M N_\text{D3}}{4\delta}\right] \,.
\end{align}
We furthermore require that
\be
V(r_*)=m= \frac{\sqrt{g_s}}{2} M N_\text{D3} \epsilon\,.
\ee
The solutions of \eqref{eq:StableShellConditions} for $\delta$ and $\epsilon$ are independent of $g_s$ and only depend on the combinations $\ell_s^2 N_\text{D3}$ and $\ell_s^{-2} M$. The physical radius, in Planck units, is then found by multiplying $\delta$ with $gQ/\sqrt{4\pi}$.
Solutions can be found numerically and we plot them below in Figure \ref{fig:r_vs_m_gs-1}.
\begin{figure}[h]
    \centering
    \includegraphics[width=\textwidth]{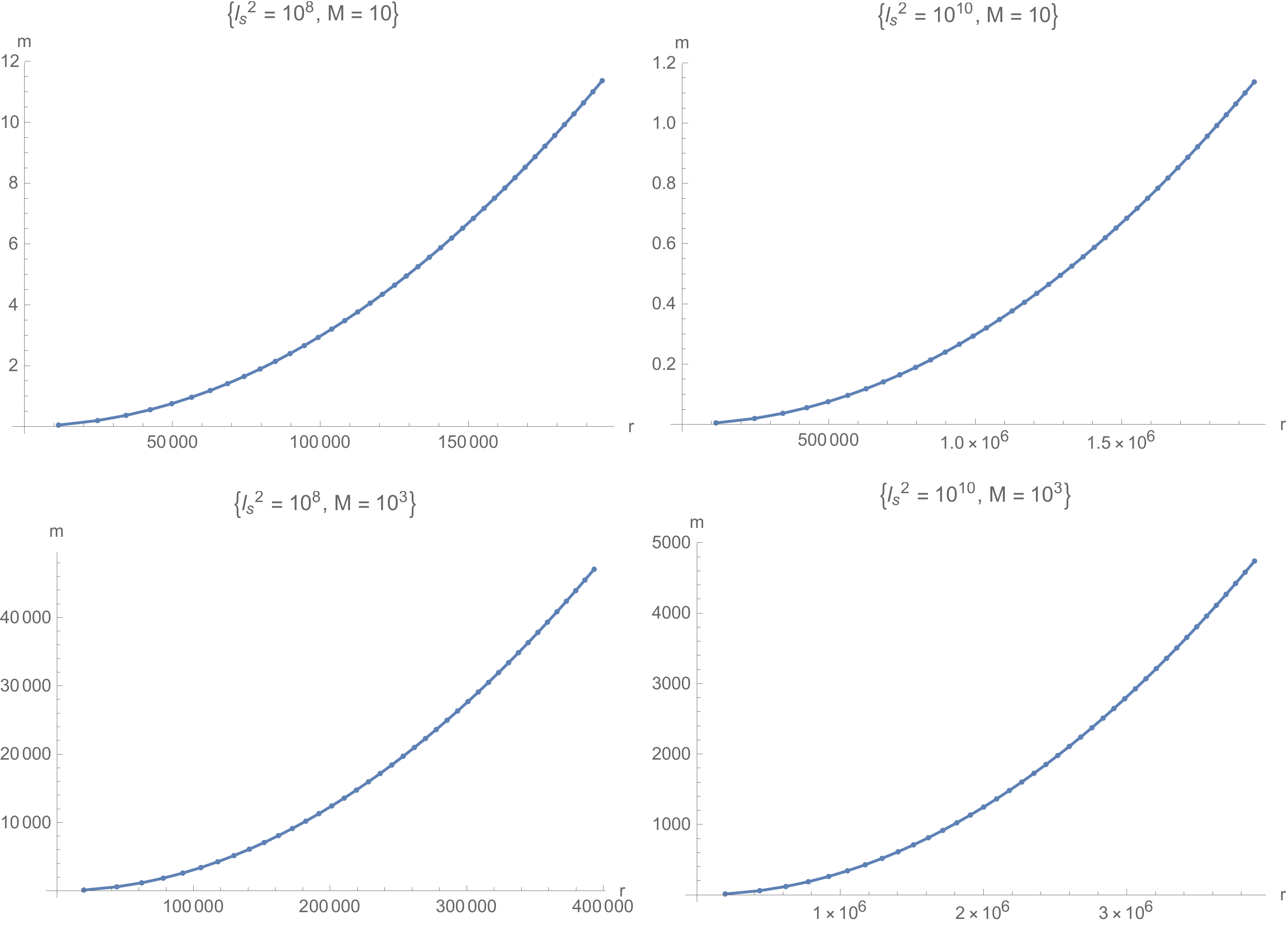}
    \caption{All these plots have $g_s=0.1$ and show the mass $m$ versus the radius $r$ for different values of the string length in Planck units and the $F_3$-flux $M$. The dots represent different values of $N_\text{D3}$, ranging from 10 from the left to 985 to the right in steps of 25 units.}
    \label{fig:r_vs_m_gs-1}
\end{figure}
Recall that the physical radius and mass our functions of $N_{\text{D3}}, M , g_s, \ell_s$ with $\ell_s$ the string length in Planck units. The last of these 3 values are ultimately determined by the compactification whereas $N_{\text{D3}}$ is chosen in the sense that it determines the state we investigate. Since that state is charged in 4d, there is no tadpole constraint, unlike for the 3-form flux quantum $M$. However, if $N_{\text{D3}}$ gets too large there are no more solutions because of the bound \eqref{taumax2}. Our numerical plots below use $g_s=0.1$, which is close to the boundary value of control, but lowering $g_s$ has a trivial effect on the physical radius and mass, as we explained before since they will just scale down as $\sqrt{g_s}$. We display for graphs with varying values for $M$ and $\ell_s$. Each of these graphs shows a range for $N_{\text{D3}}$ values between $10$ and $985$ with steps of $25$.

We observe the following general patterns
\begin{itemize}
    \item All our solutions correspond to over-extremal objects and no extremal or sub-extremal configurations can be found. This can be understood intuitively since what stabilises the shell is the Coulomb repulsion which needs to be large enough. To quantitatively see why this exactly forbids extremal or sub-extremal solutions we rely on our discussion about the maximal value of the tension-like parameter $\mathcal{\tau}$ around equation \eqref{taumax}.
    \item The larger $N_{\text{D3}}$ the larger the mass and the radius, which is to be expected.
    \item The larger $\ell_s$ the larger the physical radius and the smaller the mass, also as expected. 
    \item Less obvious are dependencies on $M$; the radius is not very sensitive to $M$, but the mass strongly increases with increasing $M$. 
\end{itemize}
Clearly all our solutions are in the regime where 10d supergravity, used to derive the configurations, can be trusted. It is interesting to contemplate how large these shells can actually become with reasonable compactification parameters $M , g_s, \ell_s$. Note that the relation between string length $\ell_s$ and Planck length $\ell_\text{Pl}$ is, up to $\pi$ factors: $\ell_s^2 =g_s^{-2}V_6 \ell_\text{Pl}^2$ with $V_6$ the compactification volume in string units. Hence, our quoted $\ell_s$ values are in the conventional ball park. 

\begin{figure}[h]
    \centering
    \includegraphics[width=0.5\textwidth]{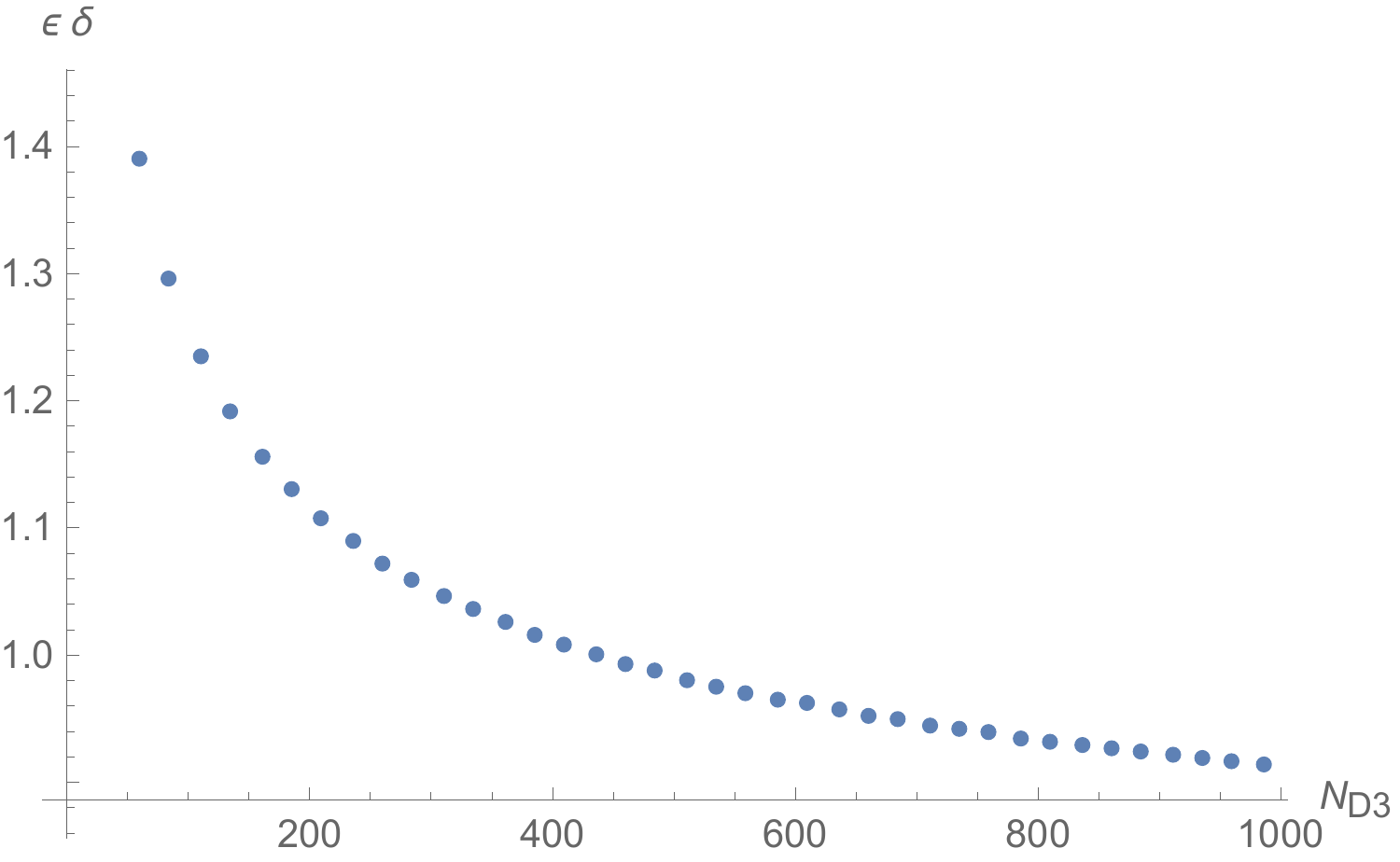}
    \caption{The product $\epsilon\delta$ for  $\ell_s^2=10^8, M=10$ and various values of $N_\text{D3}$, ranging from 10 from the left to 985 to the right in steps of 25 units.}
    \label{FIG:classicalradius}
\end{figure}

It is interesting to note that the radius we find is of the order of the ``classical radius" $r_c$ for a particle of mass $m$. The classical radius is usually defined as 
\be
r_c = \frac{g^2Q^2}{m}\,,
\ee
in natural units and up to numerical factors of order one. From our definitions above we have that
\be
r m = \frac{(gQ)^2}{4\pi} \epsilon\delta\,.
\ee
Note that the product $\epsilon\delta$ is independent of the value of $g_s$ as can be seen from minimising the shell potential. Hence, if our numerical solutions are such that the product $\epsilon\delta$ is order 1 typically we indeed find the classical radius. In figure \ref{FIG:classicalradius} we plot this product for the top-left configuration of figure \ref{fig:r_vs_m_gs-1}.
We indeed notice that all values are order unity and one can further prove analytically that the value cannot drop below $1/2$.\footnote{To see this one uses that the positive tension implies that the combination $1 - \sqrt{1 - \frac{2 \epsilon}{\delta}  + \frac{1}{\delta^2}}$ needs to be positive.}

Finally, it is useful to compare the radius with a corresponding black hole radius. Of course for over-extremal black holes there is no black hole horizon, so the best guess is to use the radius of the corresponding extremal object, which is $r_{ex}\sim \frac{gQ}{\Mpl}$ and is a good measure for the length scale associated to the gravitational backreaction. We then find that
\be
\frac{r}{r_{ex}} \approx \frac{gQ \Mpl}{m} >1\,,
\ee
because our configurations are over-extremal. So one could argue that our over-extremal fuzzball indeed corrects GR at the expected length scales.

It is interesting to compare the energy of excitations of the shell with the mass of the shell itself. Instead of solving the Schrödinger equation for perturbations, we simply use the uncertainty principle to make some estimates. For a charged shell, we have, using the junction conditions
\be
\frac{c^2 r \dot{r}^2}{2 G_4}+V(r)=m c^2 ,
\ee
with $V(r)$ our brane potential. Using the conjugate momentum, we find:
\be
\frac{G_4}{2c^2}\frac{p^2}{r}+V(r)=m c^2 .
\ee
One can estimate the level spacing by approximating the potential near the minimum as a quadratic function and then use the standard spacing formula for the harmonic oscillator. A quick and dirty way to arrive at the same formula involves Heisenberg's uncertainty relation $p r \sim \hbar$, and compare the excitation energy due the kinetic energy $\frac{G_4}{2c^2}\frac{p^2}{r}$ with $mc^2$. This relies on the fact that the quantum ground state energy is of the order of the level spacing. By using that the terms in the potential are of the same order and balancing the kinetic energy with the Coulomb energy we find that $E\sim (gQ)^3 \Mpl$. If we compare this with $mc^2$ we find that the spectrum of excitations become continuous in the limit where $m << (gQ)^3 \Mpl$, while for $m >> (gQ)^3 \Mpl$ the first excitation is much heavier than the ground state. Our solutions belong to the first category, as expected for classical configurations.

\section{Discussion}\label{sec:disc}
Let us recapitulate what we have done. We found that in the ``vanilla'' flux compactifications involving warped throats with SUSY-breaking anti-branes, D3-particles puff into spherical shells. These shells can be regarded as stabilised bubbles of vacuum decay, with the SUSY vacuum inside the bubble. This realises explicitly the ideas of black holes as brane shells put forward in \cite{Danielsson:2017riq, Danielsson:2021ruf, Danielsson:2021ykm}, albeit for over-extremal objects, so that the name black holes is somewhat a misnomer. The shells can be significantly larger than the Planck scale, and can be pushed towards the classical regime. From that perspective it is interesting that we find over-extremal configurations since the naked singularity one would expect, is not resolved in a quantum mechanical sense as for the particles of the Standard Model (whose wavelength is vastly bigger than the classical backreaction radius).  However, to get to actual astrophysical length scales one would have to lower the string scale to non-realistic values, values which is furthermore in tension with several Swampland bounds since it would mean the vacua exist at parametric weak coupling and large volume \cite{Ooguri:2018wrx, Lust:2019zwm}. However, note that our construction does not rely on de Sitter uplifts. Our computations are equally valid for uplifts of AdS to another AdS, as long as the cosmological length scales are large and the uplift energy small. 

Since these objects resolve the naked singularity of over-extremal particles they are somewhat exotic objects in the sense of describing over-extremal fuzzballs. Yet, they seem a genuine prediction of theories based on compactifications with warped throats. We also note,  in line with suggestions in \cite{Montero:2021otb}, that these objects are the WGC particles \cite{Arkani-Hamed:2006emk} for the U(1) gauge field living on anti-branes and in that sense it is reassuring that we found that we can only sustain over-extremal shells. 

It remains an interesting open problem to find a string theory embedding for black hole mimickers, in the style of \cite{Danielsson:2017riq, Danielsson:2021ruf, Danielsson:2021ykm}, using  shells that have realistic properties from an astrophysics viewpoint. 

\subsection*{Acknowledgments}
We like to thank Rob Tielemans and Miguel Montero for useful discussions. The work of TVR is supported by the KU Leuven C1 grant ZKD1118C16/16/005. VVH is supported by grant nr. 1185120N of the Research Foundation - Flanders (FWO).

\bibliographystyle{JHEP}
\bibliography{refs}

\end{document}